\documentclass[10pt,final,technote]{IEEEtran}

\usepackage[dvips]{graphicx}
\usepackage{float}
\usepackage{times}
\usepackage{amsmath}
\usepackage{amsfonts}
\renewcommand{\baselinestretch}{1.08}
\begin{document}
%
\title{A Compressive Sensing Based Approach to Sparse Wideband Array Design}

\author{\IEEEauthorblockN{Matthew B. Hawes and Wei Liu}\\
\IEEEauthorblockA{Communications Research Group\\ Department of
Electronic and Electrical Engineering\\ University of Sheffield,
UK\\{\{elp10mbh, w.liu\}@sheffield.ac.uk }}}

\maketitle

\begin{abstract}
Sparse wideband sensor array design for sensor location optimisation is highly nonlinear
and it is traditionally  solved  by genetic algorithms, simulated annealing or other similar
optimization methods. However, this is an extremely time-consuming process and more
efficient solutions are needed. In this work, this problem is studied from the viewpoint of
compressive sensing and a formulation based on a modified $l_1$ norm is derived. As there
are multiple coefficients associated with each sensor, the key is to make sure that these
coefficients are simultaneously minimized in order to discard the corresponding sensor
locations.   Design examples are provided to verify the
effectiveness of the proposed methods.
\end{abstract}

\begin{keywords}
Sparse array, frequency invariant beamforming, wideband beamforming, tapped delay-line, compressive sensing.
\end{keywords}

\section{Introduction}
Wideband beamforming has been studied extensively in the past
\cite{vantrees02a,liu2010}. It is well-known that in
order to avoid the spatial aliasing problem for uniform linear
arrays (ULAs), the adjacent sensor spacing has to be less than half
of the minimum operating wavelength corresponding to the highest
frequency of the signal of interest. On the other hand, sparse
arrays allow adjacent sensor separations greater than half a
wavelength while still avoiding grating lobes due to the randomness
of sensor locations~\cite{jarske88}.

However, the unpredictable sidelobe behaviour associated with sparse
arrays means some optimisation of sensor locations is required to
reach an acceptable performance level.  Various nonlinear methods
have been used to achieve this required optimisation.  For example,
Genetic Algorithms (GAs)
\cite{Haupt94,Yan97a,liu12n} and Simulated
Annealing (SA) \cite{Trucco99a} have been regularly used.  The
disadvantage of these types of methods are the potentially extremely
long computation times and the possibility of convergence to a
non-optimal solution.

Recently, the area of compressive sensing (CS) has been
explored \cite{Candes06}, and CS-based methods have been proposed in
the design of narrowband sparse arrays
\cite{Li08,Prisco11,Carin09,liu13g}. Further work has also shown that it is possible to improve the
sparseness of a solution by considering a reweighted $l_{1}$
minimisation problem \cite{Candes08,Fuchs12,Prisco12,liu14c}.  The aim of
these methods is to bring the minimisation of the $l_{1}$ norm of
the weight coefficients closer to that of the minimisataion of the
$l_{0}$ norm.

However, for a wideband array to be sparse, all coefficients along the tapped delay-lines (TDLs)
associated with an individual sensor have to be equal or very close to zero.  Therefore, it is not sufficient to simply minimize the $l_{1}$ norm of the weight coefficients. Instead,
a method similar to the technique employed in complex-valued $l_1$
norm minimization~\cite{Winter05}, is presented in this paper, which can be considered as an expanded version of the idea in \cite{liu13h}. As in the case with  narrowband array design, it is possible to
use a reweighted scheme for the wideband method as well. 


The remainder of this paper is structured as follows. Sec.
\ref{sec:AM} gives details of  the array model, followed by  the proposed design methods in Sec. \ref{sec:CS}.  Design examples are provided
in Sec. \ref{sec:sim} and conclusions are drawn in Sec.
\ref{sec:con}.


\section{Wideband Array Model}\label{sec:AM}

\begin{figure}
\begin{center}
   \includegraphics[angle=0,width=0.45\textwidth]{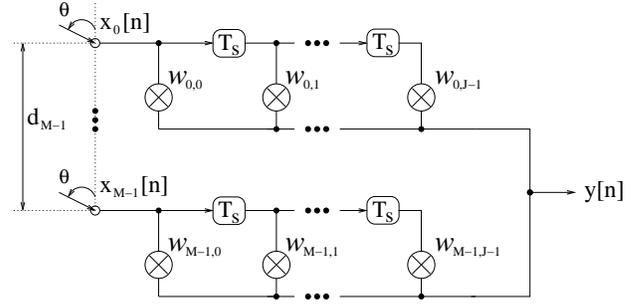}
   \caption{A general wideband beamforming structure with a TDL length $J$.
    \label{fig:AS}}
\end{center}
\end{figure}

A general linear array structure for wideband beamforming with a TDL
length J is shown in Fig. \ref{fig:AS}, where $T_{s}$ is the
sampling period or temporal delay between adjacent signal
samples~\cite{liu2010}. The beamformer output
$y[n]$ is a linear combination of differently delayed versions of
the received array signals $x_m[n]$, $m=0, \cdots, M-1$. The distance from the zeroth
sensor to the subsequent sensor is denoted by $d_{m}$ for $m=0,
\cdots, M-1$, where $d_{0} = 0$ as it is the distance from the
zeroth sensor to itself. 

The steering vector of the array as a function of the normalized frequency $\Omega=\omega T_s$ and the angle of arrival $\theta$ is
\begin{eqnarray}
\label{eq:s}\nonumber
    \textbf{s}(\Omega,\theta)&=&[1, \cdots, e^{-j\Omega(J-1)},\\ \nonumber && e^{-j\Omega\mu_{1}\cos(\theta)}, e^{-j\Omega(\mu_{1}\cos(\theta) + 1)},\\ \nonumber &&\cdots, e^{-j\Omega(\mu_{1}\cos(\theta) + (J-1))},\\ && \cdots, e^{-j\Omega(\mu_{M-1}\cos(\theta) +
    (J-1))}]^{T}.
\end{eqnarray}
where $\mu_{m}=\frac{d_{m}}{cT_{s}}$ for $m=0, 1, \cdots, M-1$ and
$\{\cdot\}^{T}$ indicates transpose operation.

The response of the array is then given by
\begin{equation}
\label{eq:response}
    P(\Omega,\theta)=\textbf{w}^{H}\textbf{s}(\Omega,\theta),
\end{equation}
where $\textbf{w}^{H}$ is the Hermitian transpose of the weight
coefficient vector of the array, given by
\begin{equation}
\label{eq:W}
    \textbf{w} = [\textbf{w}_{0}^{T}\;\; \textbf{w}_{1}^{T}
    \;\;...\;\;\textbf{w}_{M-1}^{T}]^{T}\;,
\end{equation}
with
\begin{equation}\label{eq:wm}
    \textbf{w}_{m}=[\text{w}_{m,0}\;\; \text{w}_{m,1}
    \;\;...\;\;\text{w}_{m,J-1}]^{T}.
\end{equation}

\section{Sparse Wideband Array Design via Compressive Sensing}
\label{sec:CS}
As previously mentioned, CS has been employed in the design of
sparse narrowband arrays by trying to match the array's response to
a desired/reference one, $P_{r}(\Omega,\theta)$. Extending the
design to the wideband case, we first consider Fig. \ref{fig:AS} as
being a grid of potential active sensor locations.  In this instance,
$d_{M-1}$ is the maximum aperture of the array and the values of
$d_{m}$, for $m=1, 2, \ldots, M-2$, are selected to give a uniform
grid, with $M$ being a large enough number to cover all potential
locations of the sensors. Sparseness is then introduced by selecting
the set of weight coefficients to give as few active locations as
possible, while still giving a designed response that is close to
the desired one.

In the first instance, this problem could be formulated as
\begin{eqnarray}\label{eq:min0} \nonumber
    &\min&||\textbf{w}||_{0} \\ &\text{subject to}&
    ||\textbf{p}_{r}-\textbf{w}^{H}\textbf{S}||_{2}\leq\alpha\;,
\end{eqnarray}
where $||\textbf{w}||_{0}$ is the number of nonzero weight
coefficients in $\textbf{w}$, $\textbf{p}_r$ is the vector holding
the desired beam response at sampled frequency points $\Omega_k$ and
angle $\theta_l$, $k=0, 1, \cdots, K-1$, $l=0, 1, \cdots, L-1$,
$\textbf{S}$ is the matrix composed of the steering vectors at the
corresponding frequency $\Omega_k$ and angle $\theta_l$, and
$\alpha$ places a limit on the allowed difference between the
desired and the designed responses.  In the constraint in
(\ref{eq:min0}) $||.||_{2}$ denotes the $l_{2}$ norm.

In detail, $\textbf{p}_r$ and $\textbf{S}$ are respectively given by
\begin{eqnarray}
\label{eq:pr1}\nonumber
    \textbf{p}_r&=&[P_r(\Omega_0,\theta_0), \cdots, P_r(\Omega_0,\theta_{L-1}),\\ \nonumber && P_r(\Omega_1,\theta_0), \cdots, P_r(\Omega_1,\theta_{L-1})\\ \nonumber &&...,P_r(\Omega_{K-1},\theta_{L-1})]
\end{eqnarray}
and
\begin{eqnarray}
\label{eq:S}\nonumber
    \textbf{S}&=&[\textbf{s}(\Omega_0,\theta_0), \cdots, \textbf{s}(\Omega_0,\theta_{L-1}),\\ \nonumber && \textbf{s}(\Omega_1,\theta_0), \cdots, \textbf{s}(\Omega_1,\theta_{L-1}), \cdots, \textbf{s}(\Omega_{K-1},\theta_{L-1})].
\end{eqnarray}
Here the desired response $P_{r}(\Omega,\theta)$ can be obtained
from that of a traditional uniform linear array, or simply assumed
to be an ideal response (i.e. one at the mainlobe area and zero for
the sidelobe area) and this is adopted in what follows.

In practice, the cost function in (\ref{eq:min0}) will be replaced by the $l_{1}$ norm,
\begin{eqnarray}\label{eq:min1}\nonumber
    &\min&||\textbf{w}||_{1}\\ &\text{subject to}&
    ||\textbf{p}_r-\textbf{w}^{H}\textbf{S}||_{2}\leq\alpha\;.
\end{eqnarray}

The above formulation is effective in the design of narrowband
arrays, where the TDL length $J=1$ (i.e. each sensor has only one
weight coefficient attached) and the number of nonzero coefficients
will be the same as the number of active sensors. In other words,
any coefficient with a zero value will mean that the associated
sensor is inactive.  However, in the wideband case,  to guarantee a sparse solution, the weight coefficients
along a TDL have to be simultaneously minimized.  When all
coefficients along a TDL are zero-valued, we can then consider the
corresponding location to be inactive and sparsity is introduced.
To achieve this, similar to the technique used in complex-valued
$l_1$ norm minimization~\cite{Winter05}, we minimize a modified
$l_1$ norm as follows~\cite{liu13h},
\begin{eqnarray}\label{eq:mint}\nonumber
&\min& t \;\; \epsilon\;\; \mathbb{R}^{+}\\ \nonumber&\text{subject
to}&
    ||\textbf{p}_{r}-\textbf{w}^{H}\textbf{S}||_{2}\leq\alpha\\
    &&|\langle\textbf{w}\rangle|_{1}\leq t
\end{eqnarray}
where
\begin{equation}\label{eq:constraint2}
    |\langle\textbf{w}\rangle|_{1}=\sum_{m=0}^{M-1}\Bigg|\Bigg|\left[
                                                     \begin{array}{c}
                                                       w_{m,0} \\
                                                       \vdots \\
                                                       w_{m,J-1} \\
                                                     \end{array}
                                                   \right]\Bigg|\Bigg|_{2}.
\end{equation}

Now we decompose $t$ to $t=\sum_{m=0}^{M-1}t_{m}$,
$t_{m}\epsilon\;\;\mathbb{R}^{+}$. In vector form, we have
\begin{equation}
  t=[1, \cdots, 1]\left[
                                                     \begin{array}{c}
                                                       t_{0} \\
                                                       \vdots \\
                                                       t_{M-1} \\
                                                     \end{array}
                                                   \right]=\textbf{1}^T\textbf{t}.
\end{equation}
Then (\ref{eq:mint}) can be rewritten as
\begin{eqnarray}\label{eq:min1t}\nonumber
&\min\limits_{\textbf{t}}& \textbf{1}^T\textbf{t}\\ \nonumber &\text{subject
to}&
    ||\textbf{p}_{r}-\textbf{w}^{H}\textbf{S}||_{2}\leq\alpha\\&&
    \Bigg|\Bigg|\left[
                                                     \begin{array}{c}
                                                       w_{m,0} \\
                                                       \vdots \\
                                                       w_{m,J-1} \\
                                                     \end{array}
                                                   \right]\Bigg|\Bigg|_{2}\leq t_{m},\; \;m=0, \cdots, M-1.\nonumber\\
\end{eqnarray}

Now define
\begin{equation}\label{eq:w_hat}
    \hat{\textbf{w}} = [t_{0}, w_{0,0}, \cdots, w_{0,J-1}, t_{1}, \cdots, w_{M-1,J-1}]^{T},
\end{equation}
\begin{equation}\label{eq:c_hat}
    \hat{\textbf{c}} = [1, \textbf{0}_{J}, 1, \textbf{0}_{J}, \cdots, \textbf{0}_{J}]^T
\end{equation}
and
\begin{eqnarray}
\label{eq:s_hat}\nonumber
    \hat{\textbf{s}}(\Omega,\theta)&=&[0, 1, \cdots, e^{-j\Omega(J-1)},\\ \nonumber&& 0, e^{-j\Omega\mu_{1}\cos(\theta)}, e^{-j\Omega(\mu_{1}\cos(\theta) + 1)}, \cdots,\\ \nonumber&&e^{-j\Omega(\mu_{1}\cos(\theta) + (J-1))}, \\ && \cdots, e^{-j\Omega(\mu_{M-1}\cos(\theta) + (J-1))}]^{T},
\end{eqnarray}
where $\textbf{0}_{J}$ is an all-zero  $1\times J$ row vector.
A matrix  $\hat{\textbf{S}}$ similar to (\ref{eq:S}) can be created from
$\hat{\textbf{s}}$, given by
\begin{eqnarray}
\label{eq:S_hat}\nonumber
    \hat{\textbf{S}}&=&[\hat{\textbf{s}}(\Omega_0,\theta_0), \cdots, \hat{\textbf{s}}(\Omega_0,\theta_{L-1}),\\ \nonumber && \hat{\textbf{s}}(\Omega_1,\theta_0), \cdots, \hat{\textbf{s}}(\Omega_1,\theta_{L-1}), \cdots, \hat{\textbf{s}}(\Omega_{K-1},\theta_{L-1})].
\end{eqnarray}

Finally we arrive at the final formulation for the sparse wideband
sensor array design problem
\begin{eqnarray}\label{eq:cw}\nonumber
&\min\limits_{\hat{\textbf{w}}}&
\hat{\textbf{c}}^{T}\hat{\textbf{w}} \\ \nonumber&\text{subject to}&
    ||\textbf{p}_{r}-\hat{\textbf{w}}^{H}\hat{\textbf{S}}||_{2}\leq\alpha\\&&
    \Bigg|\Bigg|\left[
                                                     \begin{array}{c}
                                                       w_{m,0} \\
                                                       \vdots \\
                                                       w_{m,J-1} \\
                                                     \end{array}
                                                   \right]\Bigg|\Bigg|_2\leq t_{m},\; \;m=0, \cdots, M-1.\nonumber\\
\end{eqnarray}

In the above formulation, it is straightforward to add additional constraint to meet some specific design requirements. For example, we can add the response variation constraint $RV=||\textbf{L}^{T}\hat{\textbf{w}}||^{2}_{2}\leq\sigma^{2}$ derived in \cite{Liu11f,Liu11d} to design a sparse wideband array with frequency invariant beam response \cite{liu07e,liu08a,liu09a,Liu11f}, where the matrix $\textbf{L}$ and the threshold value $\sigma$ are formulated to make sure the change of response of the resultant beamformer with respect to different frequencies is limited to an acceptable level.


Moreover, to increase sparsity of the resultant array, we can adopt the reweighted $l_{1}$ minimisation approach in \cite{liu14c} and reformulate \eqref{eq:cw} into the following form
\begin{eqnarray}\label{eq:rewb}\nonumber
&\min\limits_{\hat{\textbf{w}}}&
\hat{\textbf{c}}^{T}\hat{\textbf{w}} \\ \nonumber&\text{subject to}&
    ||\textbf{p}_{r}-\hat{\textbf{w}}^{H}\hat{\textbf{S}}||_{2}\leq\alpha\\&&
    a^{i}_{m}\Bigg|\Bigg|\left[
                                                     \begin{array}{c}
                                                       w^{i}_{m,0} \\
                                                       \vdots \\
                                                       w^{i}_{m,J-1} \\
                                                     \end{array}
                                                   \right]\Bigg|\Bigg|_2\leq t_{m}^{i},\nonumber\\
                                                  &&\qquad \qquad m=0, \cdots, M-1\;,
\end{eqnarray}
where
\begin{equation}\label{eq:w_hatre}
    \hat{\textbf{w}} = [t^{i}_{0}, w^{i}_{0,0}, \cdots, w^{i}_{0,J-1}, t^{i}_{1}, \cdots, w^{i}_{M-1,J-1}]^{T},
\end{equation}
\begin{equation}\label{eq:c_hatre}
    \hat{\textbf{c}} = [a^{i}_{0}, \textbf{0}_{J}, a^{i}_{1}, \textbf{0}_{J}, \cdots, \textbf{0}_{J}]^T
\end{equation}
and
\begin{equation}\label{eq:a}
    a^{i}_{m} = \Bigg(\Bigg|\Bigg|\left[\begin{array}{c}
                            w^{i-1}_{m,0} \\
                            \vdots \\
                            w^{i-1}_{m,J-1} \\
                            \end{array}
    \right]\Bigg|\Bigg|_2 + \epsilon\Bigg)^{-1}
\end{equation}
Here $\epsilon>0$ and $i$ is the iteration number.

The above problem can be solved using cvx, a package for specifying and solving convex
programs \cite{cvx,Grant08}.
\section{Design Examples}\label{sec:sim}
In this section design examples are
presented, which were all implemented on a
computer with an Intel Core Duo CPU E6750 (2.66GHz) and 4GB of RAM.
 Comparisons will  be drawn with a GA-based design method, which optimises the locations given a fixed number of sensors. In the GA based design, the fitness value was chosen to be $J_{CLS}^{-1}$, where  $J_{CLS}$ is defined in \cite{Liu11f}.

In the following example, the reference pattern was that of an ideal
array with the mainlobe at $\theta_{m}=90^{\circ}$ and sidelobe
regions of $\Theta_{s}=[0^{\circ},80^{\circ}]\bigcup[100^{\circ},180^{\circ}]$,
which were sampled every $1^{\circ}$. The frequency range
of interest $\Omega_{I}=[0.5\pi,\pi]$ was sampled every $0.05\pi$,
with the reference frequency $\Omega_{r}=\pi$. A grid of $100$
potential sensor locations was spread uniformly over an aperture of
$10\lambda$, where  the value of $\lambda$ is the wavelength
associated with a normalized frequency of $\Omega = \pi$. The values $\alpha=0.9$, $\sigma=0.01$,
$\epsilon=9\times10^{-4}$ and TDL length $J=25$ were used.

\begin{figure}
\begin{center}
   \includegraphics[angle=0,width=0.48\textwidth]{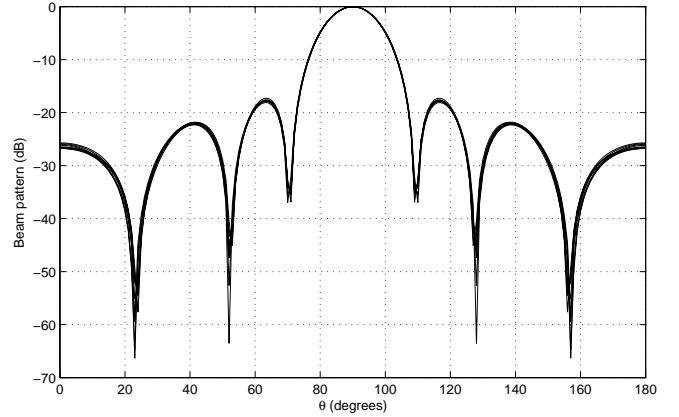}
   \caption{Responses for reweighted broadside design example.
    \label{fig:re-broad}}
\end{center}
\end{figure}

The resulting array using the reweighted method was made up of $11$ active sensor locations as
given in Tab. \ref{tb:re-broad}, with its beam response shown in
Fig. \ref{fig:re-broad}.  It can be seen that the mainlobe is at the
desired location for each normalised frequency and sufficient
attenuation has been achieved in the sidelobe regions.  The response
also shows a good level of performance in terms of the FI property.
\begin{table}
\caption{Sensor Locations For The Reweighted Broadside Design
Example.} \centering
\begin{tabular}{|c|c|c|c|}\hline
n  & $d_{n}/\lambda$ & n & $d_{n}/\lambda$   \\
\hline
0  & $1.92$ &  6  & $5.66$    \\
\hline
1  & $2.83$ &  7& $6.26$    \\
\hline
2  & $3.33$ & 8& $6.67$     \\
\hline
3  & $3.74$ & 9 & $7.17$    \\
\hline
4  & $4.34$ & 10 & $8.08$     \\
\hline
5  & $5.00$      \\
\cline{1-2}
\end{tabular}
\label{tb:re-broad}
\end{table}

\begin{figure}
\begin{center}
   \includegraphics[angle=0,width=0.48\textwidth]{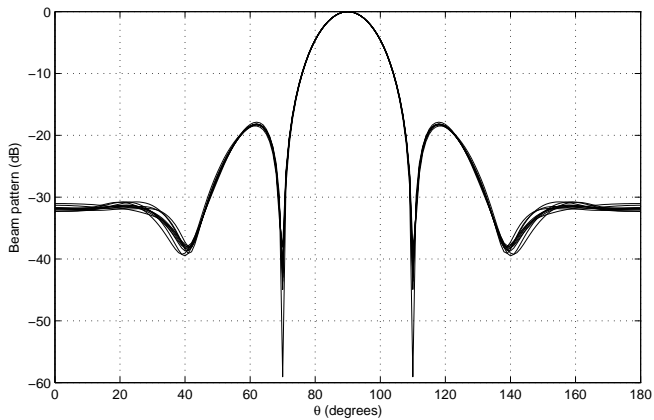}
   \caption{Responses for the GA broadside design example.
    \label{fig:broadGA}}
\end{center}
\end{figure}

This was then compared to an array designed using the GA-based
method.  To allow a fair comparison, the GA was set to optimise $11$
sensor locations over an aperture of $6.16\lambda$, the same as the example given in Tab. \ref{tb:re-broad}.  Fig. \ref{fig:broadGA} shows the resulting array response.
All these show a good performance in terms of both sidelobe attenuation and the FI property.


Tab. \ref{tb:broadcomp} summarises the different performance
measures for each design method.  The main disadvantage of the GA
design method is clearly shown, i.e. the computation time is
significantly longer.  The mean
adjacent sensor spacings are the same in both cases and larger than
the spacing of an equivalent ULA. This suggest some sparsity has
been achieved, with the same level in both cases.  Finally the value
of $|J_{CLS}|$ is slightly lower for reweighted CS designed
array, suggesting that in this instance the reweighted wideband CS
method has provided a more desirable response (the difference
largely being the FI property in the extremes of the sidelobe
regions). This will not be guaranteed to be the case all the time.

\begin{table}
\caption{Broadside Performance Comparison.} \centering
\begin{tabular}{|c|c|c|}\hline
Method                     & Reweighted & GA    \\
\hline
Mean Spacing/$\lambda$       & 0.62 &   0.62   \\
\hline
$J_{CLS}$                    & 0.0372 &  0.0376     \\
\hline
Computation Time (minutes) & 130 &   436   \\
\hline
\end{tabular}
\label{tb:broadcomp}
\end{table}

\section{Conclusions}\label{sec:con}
In this paper, a CS-based method for the design of sparse wideband arrays has been proposed, where a modified $l_1$ norm minimization problem is derived to simultaneously minimize the coefficients along a tapped delay-line associated with each sensor. Extra constraints can then be added to meet the specific design requirements, such as the frequency invariant constraint.  To further improve the sparsity of the final design result, an iterative process is employed with a reweighting term introduced in the cost function. Design examples have been presented, with comparisons also drawn with a GA-based method. Similar performance levels are achieved but the GA design takes considerably longer to reach the solution, highlighting the advantage of our proposed design methods.

\end{document}